\documentclass[twocolumn,showpacs,preprintnumbers,amsmath,amssymb,prl,aps,superscriptaddress]{revtex4}
\usepackage{graphicx}
\usepackage{dcolumn}
\usepackage{bm}
\usepackage{color}

\begin{document}
%_______________________________________________________________________________________

\title{Optomechanics of a Quantum-Degenerate Fermi Gas}

\author{R. Kanamoto}
\affiliation{Division of Advanced Sciences, Ochadai Academic Production,\\
Ochanomizu University, Bunkyo-ku, Tokyo 112-8610 Japan}

\author{P. Meystre}
\affiliation{B2 Institute, Department of Physics and College of Optical Sciences,\\
The University of Arizona, Tucson, Arizona 85721 USA}

\date{\today}

%_______________________________________________________________________________________

\begin{abstract}
We explore theoretically the optomechanical interaction between a light field and a mechanical mode of ultracold fermionic atoms inside a Fabry-P\'{e}rot cavity. The low-lying phonon mode of the fermionic ensemble is a collective density oscillation associated with particle-hole excitations, and is mathematically analogous to the momentum side-mode excitations of a bosonic condensate. The mechanical motion of the fermionic particle-hole system behaves hence as a ``moving mirror.'' We derive an effective system Hamiltonian that has the form of generic optomechanical systems. We also discuss the experimental consequences the optomechanical coupling in optical bistability and in the noise spectrum of the system.
\end{abstract}
\maketitle
%_______________________________________________________________________________________

Optomechanics, the study of the mechanical effects of light on mesoscopic and macroscopic mechanical oscillators, is an emerging field with considerable promise for applications ranging from quantum metrology at or beyond the standard quantum limit~\cite{03:CHP,06:ABHP}, and from fundamental studies of quantum mechanics, including the cooling of macroscopic objects to their quantum mechanical ground state~\cite{07:KV} to the study of the quantum-classical interface~\cite{02:Legg}, and the generation of quantum superposition of macroscopic objects~~\cite{99:BJK,02:MGVT,03:MSPB}.
The cornerstone of these developments is the control of mechanical degrees of freedom by optical forces, most prominently radiation pressure. Optomechanical systems of particular interest include optical resonators with a movable end-mirror~\cite{83:DMMVW}, membranes or mirrors placed inside an optical cavity~\cite{84:MMW,85:MWMV,08:Yale}, and nanoscale cantilevers~\cite{08:MPI}.

In complementary work, recent experiments have exploited the coupling of the center-of-mass degrees of freedom of ultracold atomic ensembles to the optical field inside a cavity to realize optomechanical systems. In such experiments an analog of a moving mirror can be provided, e.g., by side-mode excitations of a bosonic condensate~\cite{08:BRDE, 09:GZM} and by the vibrational motion of cold thermal atoms~\cite{07:GMMS, 08:MMGS}. In a sense, these experiments can be thought of as "bottom up" realizations of optomechanical systems, in contrast to the "top down" systems realized by mirrors, membranes or cantilevers. Here, photon recoil results in the excitation of atomic density fluctuations associated with the occupation of momentum side modes. Experimental signatures of this effect include the onset of optical bistability in the transmitted light~\cite{83:DMMVW}.

This Letter extends these ideas to ultracold {\it fermionic} gases, and show how a mechanical mode of ultracold spinless fermionic atoms trapped inside an optical resonator can couple optomechanically to the cavity field. In the limit of low photon numbers, or equivalently the regime where only the lowest diffraction order of the atoms by the light fields is significantly excited, we identify a complete analogy between a mechanical fermionic mode consisting of a superposition of particle-hole excitations and the more familiar momentum side mode excitations of a bosonic condensate. Evidence of the optomechanical coupling can again be observed in the bistable behavior in the semiclassical stationary state of the system, as well as in its noise spectrum.

The quantum statistics of atoms are often irrelevant in cavity QED experiments
using atomic ensembles. However, the situation is different when dealing with the center-of-mass degrees of freedom of ultracold atoms. In this regime of cavity QED and atom optics experiments, bosonic condensates are often preferred due to their formal similarity with photons, and also because one often wants to exploit Bose enhancement. On the other hand, condensate dynamics is often dominated by atomic collisions and collisional frequency shifts that can mask or even suppress the effects of atom-field interactions. This is one reason for the interest of fermions instead of bosons in high-performance atomic
clocks~\cite{95:GV,03:Kat} and interferometers~\cite{04:MI}, see also Refs.~\cite{01:KI,01:MM,03:CSM,05:MSM,99:RCB,08:LML} which consider some aspects of fermionic atom optics.

% ----------------------------
\begin{figure}
\includegraphics[scale=0.45]{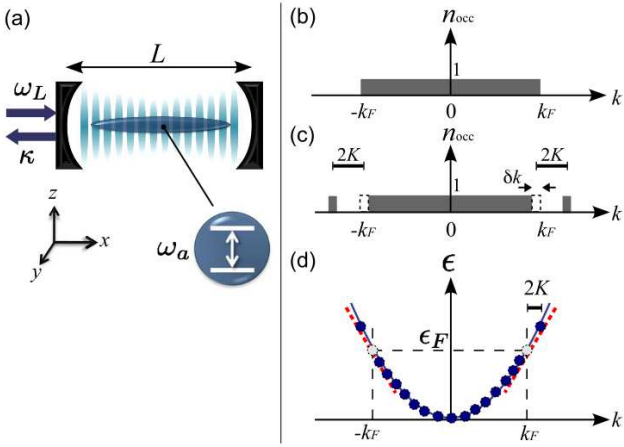}
\caption{(Color online) (a) A sample of two-level fermionic atoms with resonant frequency $\omega_a$ is trapped in a Fabry-P\'{e}rot cavity of length $L$, and interact with the standing-wave light field. The left mirror is partially transmissive and the cavity is driven by a laser of frequency $\omega_L$.
(b) Ground-state atomic momentum distribution in the absence of grating, with $k_F$ the Fermi momentum.(c) Atomic momentum distribution resulting from the diffraction of atoms off the grating for low intracavity photon numbers. States within the width $\delta k$ close to $k_F$ acquire a momentum boost $2K$ due to photon recoil, with a conjugate process close to $-k_F$ with momentum boost $-2K$.
(d) Energy dispersion corresponding to (c). For $2K \ll k_F$ the quadratic dispersion relation of the atoms can be approximated by its gradient at the Fermi energy.
}
\label{fig1}
\end{figure}
% ----------------------------

Figure~\ref{fig1}(a) shows the system under consideration. A Fabry-P\'{e}rot cavity of length $L$ containing $N$ trapped spinless fermionic atoms of mass $M$ is driven at rate $\eta$ by a pump laser of frequency $\omega_L$ and wave number $K$. We assume that the magnetic trapping frequency is much tighter in the transverse $y$ and $z$ directions than along the cavity axis $x$
so that low-energy excitations of fermions occur predominantly occur along $x$. We quantize the atomic motional degree of freedom along that axis and describe its dynamics within a one-dimensional model.

When the pump laser frequency $\omega_L$ is far detuned from the atomic
transition frequency $\omega_a$, the excited electronic state of the atoms can be adiabatically eliminated and the atoms interact dispersively with the cavity field, taken to be single mode. In the dipole and rotating-wave approximations, the atomic part of Hamiltonian is then~\cite{05:MR}
\begin{equation}
\int dx \hat{\Psi}^{\dagger}(x)\left[\hat{p}_x^2/(2M)+\hbar U_0\cos^2 (Kx)\hat{c}^{\dagger}\hat{c}\right]\hat{\Psi}(x),
\end{equation}
where $\hat{\Psi}(x)$ is the atomic field operator, and
$\hat{c}$ the annihilation operator of a cavity photon. Because of the absence of low-energy $s$-wave collisions between identical fermions, they interact
only with the light field with the coupling $U_0=g_0^2/(\omega_L-\omega_a)$, where $g_0$ is the single-photon Rabi frequency.

We expand the atomic field operator in terms of plane waves
$\hat{\Psi}(x)=L^{-1/2}\sum_k \hat{f}_k e^{ikx}$ where $\hat{f}_k$ is a
annihilation operator, with
$\{\hat{f}_k,\hat{f}^{\dagger}_{k'}\}=\delta_{k,k'}$, $\{\hat{f}_k,\hat{f}_{k'}\}=0$.
This yields the second-quantized Hamiltonian~\cite{03:CSM,05:MSM,99:RCB}
\begin{eqnarray}\label{Hf}
\hat{H}&=&\sum_k \epsilon (k)\hat{f}_k^{\dagger}\hat{f}_k
+\hbar \Delta\hat{c}^{\dagger}\hat{c}\nonumber\\
&&\qquad +\frac{1}{4}\hbar U_0 \hat{c}^{\dagger}\hat{c}
\sum_k(\hat{f}_{k+2K}^{\dagger}\hat{f}_k+\hat{f}_k^{\dagger}\hat{f}_{k+2K})
\end{eqnarray}
where $\epsilon(k)=\hbar^2k^2/(2M)$ is the kinetic energy of a fermion of momentum $k$, $\Delta=\omega_c-\omega_L+U_0N/2$ is the effective cavity detuning,
and $\omega_c$ the resonant frequency of the empty cavity nearest to the laser frequency.

Because of photon recoil by the off-resonant standing-wave field, the atoms can suffer a momentum kick $\pm 2\ell K$, $\ell$ being the diffraction order. For feeble optical fields it is sufficient to consider the lowest diffraction order, $\ell = 1$, in which case the physics of the problem is best captured by introducing the operators
\begin{eqnarray}\label{rho}
\hat{\rho}_{p}=\sum_{k} \hat{f}_k^{\dagger}\hat{f}_{k+p},
\end{eqnarray}
where $p=2K$. It is easily verified that $\rho_{p}^{\dagger}=\rho_{-p}$.
Equation~(\ref{rho}) describes a superposition of particle-hole excitations with a well-defined excitation momentum $p$, assumed to be positive without loss of generality. We further define ``right-'' and ``left-'' propagating atomic density-fluctuation operators as $\rho_{p}^{(+)}$, $\rho_{p}^{(-)}$, corresponding to the summation of $k$ in Eq.~(\ref{rho})
for $k>0$, and $k<0$, respectively~\cite{Gia}. The examination of the commutation relations of $\hat{\rho}_{\pm p}^{(\pm)}$ leads naturally to the introduction of a new set of operators
\begin{eqnarray}
&&\hat{b}_{p}= \beta_{p}\hat{\rho}^{(+)}_{p},\quad \hat{b}_{p}^{\dagger}=\beta_{p} \hat{\rho}^{(+)}_{-p},
\nonumber\\
&&\hat{b}_{-p}=\beta_{p}\hat{\rho}^{(-)}_{-p},\quad \hat{b}^{\dagger}_{-p}=\beta_{p}\hat{\rho}^{(-)}_{p},
\end{eqnarray}
where $\beta_{p}=\sqrt{2\pi/pL}=\sqrt{\pi/KL}$ is a normalization constant. These operators obey the bosonic commutation relation
$[\hat{b}_{\pm p},\hat{b}_{\pm p}^{\dagger}]=1$, and
$[\hat{b}_{\pm p},\hat{b}_{\pm p}]=[\hat{b}_{\pm p},\hat{b}_{\mp p}^{\dagger}]=0$.
The components with opposite sign thus behave as if describing distinguishable oscillators.

We consider the simple situation where the intracavity field can be approximated as a plane wave, so that only momentum transfer along the cavity axis is considered and an effective one-dimensional description is appropriate. We also assume that the fermions are perfectly degenerate and occupy all momentum states with $k \in [-k_F,k_F]$ with interval $2\pi/L$ [Fig.~\ref{fig1}(b)], and denote by $\delta k$ the width of momentum states that can be diffracted by the optical field. Clearly that width is at most $2 \ell K$, or $2K$ for $\ell = 1$. The excitation energies of the particle-hole excitations at the edges of the interval $[k_F-\delta k,k_F]$ are $\epsilon(k_F-\delta k+2K)-\epsilon(k_F-\delta k)=2 \hbar^2 K(K+k_F-\delta k)/M $, and $\epsilon(k_F+2K)-\epsilon(k_F)=2 \hbar^2 K(K+k_F)/M$, respectively.

We assume in the following that the number of atoms is large enough that $K <  k_F=\pi N/L$,  the opposite of the regime $k_F \ll 2K$ of Ref.~\cite{05:MSM}.
In that case we can approximate the quadratic energy dispersion by its lowest order expansion about the Fermi energy, $\sum_k \epsilon (k)\hat{f}_k^{\dagger}\hat{f}_k \simeq \sum_k v_F \hbar |k|\hat{f}_k^{\dagger}\hat{f}_k$, where $v_F=\hbar k_F/M$ is the Fermi velocity, see Fig.~\ref{fig1}(d). This amounts to neglecting the dependence on $k$ of the excitation energies of the particle-hole pairs, $K(k_F\pm K)\simeq Kk_F$. From the commutation relations of $\hat{b}$ operators and
$\sum_k \hbar v_F|k|\hat{f}_k^{\dagger}\hat{f}_k$, one can then rewrite as
\begin{eqnarray}
\sum_k \epsilon(k)\hat{f}_k^{\dagger}\hat{f}_k
\to
\sum_{p>0}\hbar v_F p (\hat{b}_p^{\dagger}\hat{b}_p+\hat{b}_{-p}^{\dagger}\hat{b}_{-p}),
\end{eqnarray}
and the final form of the effective Hamiltonian is
\begin{widetext}
\begin{eqnarray}\label{Heff}
\hat{H}_{\rm eff}=\hbar \omega_M(\hat{b}_{2K}^{\dagger}\hat{b}_{2K}+\hat{b}_{-2K}^{\dagger}\hat{b}_{-2K})
+\hbar \left[\Delta +g(\hat{b}_{2K}^{\dagger}+\hat{b}_{2K})+g(\hat{b}_{-2K}^{\dagger}+\hat{b}_{-2K})\right]
\hat{c}^{\dagger}\hat{c}+i\hbar \eta(\hat{c}^{\dagger}-\hat{c}).
\end{eqnarray}
\end{widetext}
This Hamiltonian bears a marked similarity to the generic Hamiltonian of cavity optomechanics, with mechanical oscillator frequency $\omega_M=2Kv_F$, and optomechanical coupling $g=U_0/(4\beta_{2K})$. It is also formally identical to the effective Hamiltonian that describes optomechanics experiments using a bosonic condensate~\cite{08:BRDE}, except that now it is the {\it collective density fluctuation of the fermions} that plays a role of a moving mirror.
In contrast to the situation with a Bose condensate, in the fermionic system the noninteracting ground state is a filled Fermi sea. The lowest momentum side mode corresponds then to the process $|k|\to |k+2K|$ rather than $0 \to |2K|$. A key observation, though, is that although the original statistics of the atoms is fermionic, the excitation is bosonic, due to the fact that a single photon is associated with each specific particle-hole pair.

%_______________________________________________________________________________________

We next show that there exists an experimentally achievable set of parameters that satisfies the conditions invoked in deriving the effective Hamiltonian.
To guide this discussion we concentrate first on the bistable behavior in the system's steady state, one of the hallmarks of optomechanical coupling.
We proceed by introducing the quadratures of the mechanical oscillators
$\hat{X}_{M}=\hat{X}_++\hat{X}_-$, $\hat{P}_{M}=\hat{P}_++\hat{P}_-$ where
$\hat{X}_{\pm}=(\hat{b}^{\dagger}_{\pm2K}+\hat{b}_{\pm 2K})/\sqrt{2}$,
$\hat{P}_{\pm}=i(\hat{b}_{\pm 2K}^{\dagger}-\hat{b}_{\pm 2K})/\sqrt{2}$, respectively.
The corresponding Heisenberg-Langevin equations are obtained from Eq.~(\ref{Heff}) as
\begin{eqnarray}
&&\frac{d\hat{X}_M}{dt}=\omega_M\hat{P}_M,\quad
\frac{d\hat{P}_M}{dt}=-\omega_M\hat{X}_M-2\sqrt{2}g \hat{c}^{\dagger}\hat{c},\nonumber\\
&&\frac{d\hat{c}}{dt}=-i\left[\Delta +\sqrt{2}g\hat{X}_M\right]\hat{c}+\eta-\kappa \hat{c}+\sqrt{2\kappa}\hat{c}_{\rm in},
\label{langevin}
\end{eqnarray}
where $\kappa$ is the cavity decay rate and $\hat{c}_{\rm in}$ denotes a Markovian noise operator~\cite{04:GZ} associated with a reservoir at zero temperature responsible for the cavity decay. It has zero mean
and correlations $\langle \hat{c}_{\rm in}^{\dagger}(t)\hat{c}_{\rm in}(t')\rangle=2\kappa\delta(t-t')$, and
$\langle \hat{c}_{\rm in}(t)\hat{c}_{\rm in}(t')\rangle=0$. In contrast with the situation in optomechanics with real mirrors, thermal damping
can be ignored here.

% ----------------------------
\begin{figure}[b]
\includegraphics[scale=0.5]{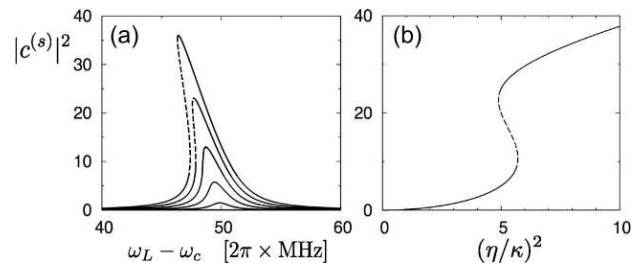}
\caption{(a) Steady-state intracavity photon number as a function of
(a) pump-cavity detuning for -- by increasing maxima --
$\eta/\kappa=$ 1.2, 2.4, 3.6 (bistability threshold), 4.8, and 6.0; (b) pump rate for $\omega_L-\omega_c=2\pi \times 47.5$ MHz.
}
\label{fig2}
\end{figure}
% ----------------------------

The mean-field steady-state solution ${\cal O}^{(s)}$ of the Langevin equations (\ref{langevin}) is readily obtained by setting all time derivatives be zero. This gives $P_M^{(s)}=0$,
$X_M^{(s)}=-2\sqrt{2}g|c^{(s)}|^2/\omega_M$, and
\begin{eqnarray}
c^{(s)}=\frac{\eta}{\kappa+i\left(\Delta-4g^2\omega_M^{-1}|c^{(s)}|^2\right)}.
\end{eqnarray}
This form of cubic equation is characteristic of optical multistability~\cite{84:MMW, 07:GMMS, 08:BRDE, 08:LML}.
We consider for illustration the experimentally achievable parameters $U_0=2\pi \times 20$ kHz, $\lambda = 500$ nm ($K\simeq 10^7$ m${}^{-1}$), $L=100$ $\mu$m, $N\simeq 5000$ atoms yielding a Fermi momentum $k_F\simeq 10^8$ m${}^{-1}$, so that $k_F=12.5K$, $\kappa=2\pi \times 1$ MHz, $M=1.5\times 10^{-25}$ kg, and the Fermi frequency $\omega_F\equiv \epsilon_F/\hbar \simeq 10$ MHz.
This value of $U_0$ assumes a single-photon Rabi frequency $g_0=2\pi \times 10$ MHz and the pump-atom detuning $\omega_L-\omega_a=2\pi \times 30$ GHz, similarly to Ref~\cite{08:BRDE}. Since the decay rate of the atomic excited state is of the order of MHz, spontaneous emission can safely be neglected here. Typical frequencies of the side-mode excitations are then
$\hbar (\delta k)^2/(2M) \simeq \hbar (2K)^2/(2M) \simeq 0.1$ MHz, so that
the assumptions invoked in deriving the effective Hamiltonian are satisfied.

Figure~\ref{fig2} shows the steady-state mean photon number as a function of
the pump-cavity detuning (a) for fixed values of $\eta/\kappa$, and (b) as a function of the pump rate for a fixed value of the pump-cavity detuning.
From the linear stability analysis described below, the steady state is easily seen to be bistable. The fluctuations around the steady state can be studied via a linearization of the Langevin equations about their steady-state mean. We proceed by expanding each operator around its steady-state value as $\hat{\cal O}(t)={\cal O}^{(s)}+\delta\hat{\cal O}(t)$, and introduce the cavity-field
quadratures $\delta \hat{X} = (\delta \hat{c}^{\dagger}+\delta \hat{c})/\sqrt{2}$, $\delta \hat{P}=i(\delta \hat{c}^{\dagger}-\delta \hat{c})/\sqrt{2}$. The linearized Langevin equations are then written in the usual form $\dot{f}(t)=J f(t)+\xi(t)$, where
$f(t)=[\delta \hat{X}_M, \delta \hat{P}_M,\delta \hat{X}, \delta \hat{P}]^T$,
the noise term $\xi(t)=[0,0,\sqrt{2\kappa}\delta \hat{X}_{\rm in},\sqrt{2\kappa}\delta \hat{P}_{\rm in}]^T$, and the drift matrix $J$ is given by
\begin{eqnarray}\label{matJ}
J=\left[
\begin{array}{cccc}
0 & \omega_M & 0 & 0 \\
-\omega_M & 0 & -4gc_s & 0\\
0 & 0 & -\kappa & \tilde{\Delta}\\
-2gc_s & 0 & -\tilde{\Delta} &-\kappa
\end{array}.
\right]
\end{eqnarray}
Here we have taken $c^{(s)}$ to be a real number without loss of generality, and $\tilde{\Delta}=\Delta-4g^2 \omega_M^{-1}c_s^2$ is the effective detuning.

The eigenvalues of the matrix~(\ref{matJ}) are obtained by direct diagonalization, and we find that the steady state is stable only if the real part of the eigenvalues is zero or negative, and as expected, the branch of the bistability curve with negative slope (dashed part in Fig.~2) is found to be unstable. Further, the imaginary part of eigenvalue corresponds to the oscillation
frequency of each quadratures. At the crest of the bistability curve, the imaginary part becomes zero.

The atom-cavity system is typically probed via measurements on the optical field transmitted by the cavity, in particular its spectrum. The noise spectrum of an arbitrary stationary field with fluctuations $\delta \hat{f}[t]$ is defined by the Fourier transform of its autocorrelation function $\langle \delta\hat{f}[\omega]\delta \hat{f}[\omega']\rangle=S_{f}[\omega]\delta(\omega+\omega')$
as
$$
S_f[\omega]=\frac{1}{2\pi}\left\{\int d\omega' e^{-i(\omega+\omega')t}
\langle \delta \hat{f}[\omega]\delta \hat{f}[\omega']\rangle\right\}_t,
$$
where $\delta \hat{f}[\omega]$ is the transform of $\delta \hat{f}[t]$
and $\{\cdots\}_t$ denotes a time average. With
$\delta X_{\rm in}=(\delta\hat{c}_{\rm in}^{\dagger}+\delta\hat{c}_{\rm in})/\sqrt{2}$ and $\delta P_{\rm in}=i(\delta\hat{c}_{\rm in}^{\dagger}-\delta\hat{c}_{\rm in})/\sqrt{2}$, the solution of the linearized Langevin equations gives
\begin{eqnarray}
&&\delta X_M [\omega]=\frac{4\sqrt{2\kappa} g c^{(s)} \omega_M
\{(\kappa-i\omega)\delta X_{\rm in}[\omega]+\tilde{\Delta}\delta P_{\rm in}[\omega]\}}
{d[\omega]},
\nonumber\\
\end{eqnarray}
where
$d[\omega]=(\omega^2-\omega_M^2)[(\kappa-i\omega)^2+\tilde{\Delta}^2]+2\omega_M\tilde{\Delta}(2gc^{(s)})^2$,
so that
\begin{equation}
S_{X_M}=\frac{2\kappa(4gc^{(s)}\omega_M)^2
\left[\kappa^2+(\tilde{\Delta}+\omega)^2\right]}
{|d[\omega]|^2}
\end{equation}
and the measurable quadratures of the optical field \cite{06:PGKBBA} are
\begin{eqnarray}
&&S_{X_c}=\frac{(2gc_s)^2\tilde{\Delta}^2S_{X_M}
+2\kappa\left[\kappa^2+(\omega+\tilde{\Delta})^2\right]
}{|d[\omega]|^2}
\\
&&S_{P_c}=\frac{(2gc_s)^2 (\kappa^2+\omega^2)S_{X_M}
+2\kappa\left[\kappa^2+(\omega+\tilde{\Delta})^2\right]
}{|d[\omega]|^2}.\nonumber
\end{eqnarray}

In conclusion, we have discussed theoretically the possibility of realizing optomechanical systems using a degenerate gas of spinless fermions. We found that a mechanical fermionic mode analogous to the conventional moving mirror of optomechanics is provided by the collective density fluctuations associated with particle-hole excitations, and this mode can be quantized as a bosonic excitation. We have also identified experimentally realistic parameters such that this effective fermionic mirror results in a bistable steady-state behavior of the intracavity photon number. Generally speaking, optomechanics with fermionic mirrors should be observable under conditions similar to the experiments using bosonic condensates~\cite{08:BRDE,09:GZM}.

Future studies will include a self-consistent determination of the fermionic state in quantized light fields, transverse effects, higher-order atomic diffraction, as well as the role of the trapping potential, which modifies the mean atom number at each site of the optical potential and may lead to parametric heating.

We acknowledge M.~Bhattacharya, W.~Chen, D.~Goldbaum, and S.~Singh for useful discussions. This work is supported in part by the National Science Foundation, the US Army Research Office, the US Office of Naval Research, and by a Japan Grant-in-Aid for Scientific Research under Grant No. 21710098.\

%_______________________________________________________________________________________

%_______________________________________________________________________________________

\end{document}